\documentclass[twocolumn,aps,superscriptaddress,showpacs]{revtex4}

\usepackage{amssymb}
\usepackage{amsmath}
\usepackage{graphicx}
\usepackage[normalem]{ulem}
\usepackage[dvips]{color}

\begin{document}
\title{Thermal properties of asymmetric nuclear matter with an improved
isospin- and momentum-dependent interaction }

\author{Jun Xu}
\email{xujun@sinap.ac.cn} \affiliation{Shanghai Institute of Applied
Physics, Chinese Academy of Sciences, Shanghai 201800, China}
\affiliation{Kavli Institute for Theoretical Physics, Chinese
Academy of Sciences, Beijing 100190, China}

\author{Lie-Wen Chen}
\email{lwchen@sjtu.edu.cn} \affiliation{Department of Physics and
Astronomy and Shanghai Key Laboratory for Particle Physics and
Cosmology, Shanghai Jiao Tong University, Shanghai 200240, China}
\affiliation{Center of Theoretical Nuclear Physics, National
Laboratory of Heavy Ion Accelerator, Lanzhou 730000, China}

\author{Bao-An Li}
\email{bao-an.li@tamuc.edu} \affiliation{Department of Physics and
Astronomy, Texas A$\&$M University-Commerce, Commerce, TX
75429-3011, USA} \affiliation{Department of Applied Physics, Xi'an
Jiao Tong University, Xi'an 710049, China}

\begin{abstract}
Thermal properties of asymmetric nuclear matter, including the
temperature dependence of the symmetry energy, single-particle
properties, and differential isospin fractionation, are investigated
with different neutron-proton effective mass splittings using an
improved isospin- and momentum-dependent interaction. In this
improved interaction, the momentum-dependence of the isoscalar
single-particle potential at saturation density is well fitted to
that extracted from optical model analyses of proton-nucleus
scattering data up to nucleon kinetic energy of 1 GeV, and the
isovector properties, i.e., the slope of the nuclear symmetry
energy, the momentum-dependence of the symmetry potential, and the
symmetry energy at saturation density can be flexibly adjusted via
three parameters $x$, $y$, and $z$, respectively. Our results
indicate that the nucleon phase-space distribution in equilibrium,
the temperature dependence of the symmetry energy, and the
differential isospin fractionation can be significantly affected by
the isospin splitting of nucleon effective mass.
\end{abstract}


\maketitle

\section{Introduction}

Understanding the in-medium nucleon-nucleon (NN) interaction is one
of the main tasks of nuclear physics. The single-particle potential
of a nucleon in nuclear medium is closely related to the NN
interaction as well as the properties of nuclear matter. Based on
the Brueckner theory, the potential of a nucleon depends not only on
the properties of the medium but also on the momentum of the
nucleon, and the momentum dependence comes from the exchange
contribution of the finite-range NN interaction within Hartree-Fock
framework. More than twenty years ago, for studying heavy-ion collisions
the momentum-dependent mean-field potential was gradually improved from the
Gale-Bertsch-Das Gupta (GBD) interaction~\cite{GBD87} to a
momentum-dependent Yukawa interaction (MDYI)~\cite{Wel88,Gal90}.
Later, the isospin-dependence was further introduced to the
momentum-dependent potential and the newly developed interaction is
named as MDI~\cite{Das03}. It has been found that the momentum
dependence of the nucleon potential affects not only the dynamics of
heavy-ion collisions (see Ref.~\cite{Li08} for a review) but the
thermodynamical properties of nuclear matter as well~\cite{Mis93,Xu08}.
This interaction has further been used to study the core-crust
transition density of neutron
stars~\cite{Nstarcrust1,Nstarcrust2,Nstarcrust3} and study the
properties of hybrid stars after it was extended to include hyperon
interactions~\cite{Xu10}. Moreover, the MDI interaction together
with an isospin-dependent Boltzmann-Uehing-Uhlenbeck transport model
was used to study the symmetry energy at both
subsaturation~\cite{Che05} and suprasaturation
densities~\cite{Xia09}. For a latest review of the MDI interaction,
we refer the readers to Ref.~\cite{Che14}.

The above MDI interaction was further improved in 2010~\cite{ImMDI},
and the new interaction, dubbed ImMDI, mainly includes the following
three improvements. First, the single-particle potential in
symmetric nuclear matter at $\rho_0$ was refitted to reproduce the
empirical optical potential by Hama {\it et al.}~\cite{Hama90,Coo93}
up to nucleon kinetic energy of 1 GeV, while that in the previous
MDI interaction becomes more attractive than that extracted from the
proton-nucleus scattering data at nucleon momenta larger than about
$550$ MeV/c (i.e., the nucleon kinetic energy of about $160$ MeV),
as can be seen from Fig.~2 of Ref.~\cite{Xu10}. Second, a parameter
$y$ was introduced to mimic the momentum dependence of the symmetry
potential, or equivalently, the isospin splitting of the nucleon
effective mass. Third, considering that the isospin tracers are
sensitive to both the slope parameter $L$ of the symmetry energy
(mimiced by the parameter $x$ in the MDI interaction) and the
symmetry energy $E_{sym}(\rho_0)$ at saturation density and the
constraints of the nuclear symmetry energy are usually mapped in the
$L \sim E_{sym}(\rho_0)$ plane (see, e.g., Fig.~1 of
Ref.~\cite{Che12} and Fig.~2 of Ref.~\cite{Tsa12}), a parameter $z$
is introduced to vary the value of $E_{sym}(\rho_0)$. The ImMDI
interaction can thus describe more reliably the dynamics of
heavy-ion collisions at beam energies up to 1 GeV and provide
possibilities to study simultaneously more detailed isovector
properties of nuclear matter, such as the slope parameter of the
symmetry energy, the momentum dependence of the symmetry potential,
and the symmetry energy at saturation density.

The neutron-proton effective mass splitting has been studied for a
long time~\cite{Liu02,ZYM04,Li04,LiBA04,Riz05} and recently becomes
again a hot
topic~\cite{Gio10,OLi11,Beh11,Fen11,Fen12,Li13,XHL13,Li14,Zha14,Xie14}.
It is noteworthy that in relativistic models one needs to calculate
the Lorentz mass so that it can be compared with that from the
non-relativistic interactions. For Lorentz effective mass, the
microscopic Brueckner-Hartree-Fock or Dirac-Brueckner-Hartree-Fock
approach and most Skyrme-Hartree-Fock calculations lead to a larger
neutron effective mass than proton in neutron-rich nuclear matter,
while most relativistic mean-field models and a few
Skyrme-Hartree-Fock calculations give opposite predictions. The
larger neutron effective mass than proton requires that the nuclear
symmetry potential decreases with increasing nucleon
momentum/energy, which is more consistent with the Lane potential in
trend~\cite{Li04}. In addition, the neutron clearly has a larger
effective mass than the proton in neutron-rich matter based on
optical model analyses for nucleon-nucleus elastic
scatterings~\cite{XuC10,XHL13,Li14}. On the other hand, the recent
experimental data of double neutron/proton ratio from the National
Superconducting Cyclotron Laboratory seems to favor a smaller
neutron effective mass than proton based on the calculation using an
improved quantum molecular dynamics model~\cite{Cou14}, although the
short-range correlation might be another alternative
explanation~\cite{Hen14}. Since the possibility of a smaller neutron
effective mass than proton in neutron-rich matter has not been
absolutely ruled out yet and is currently hotly debated, it is thus
of great interest to study in more details the possible effects from
different neutron-proton effective mass splittings. It has been
found that the dynamic properties in heavy-ion collisions can be
affected by the isospin splitting of nucleon effective mass and the
latter has considerable effects on the single and double
neutron/proton ratio, t/$^3$He ratio, and isospin-dependent
collective flows and particle
productions~\cite{LiBA04,Riz05,Gio10,OLi11,Fen11,Fen12,Zha14,Xie14}.
In the present manuscript, we will study the effects on
thermodynamical properties of nuclear matter from different isospin
splittings of nucleon effective mass based the ImMDI interaction.

\section{The improved isospin- and momentum-dependent interaction}

The functional form of potential energy density of nuclear matter
for the ImMDI interaction is the same as the MDI
interaction~\cite{Das03,Che05}, i.e.,
\begin{eqnarray}
V(\rho ,\delta ) &=&\frac{A_{u}\rho _{n}\rho _{p}}{\rho _{0}}+\frac{A_{l}}{%
2\rho _{0}}(\rho _{n}^{2}+\rho _{p}^{2})+\frac{B}{\sigma +1}\frac{\rho
^{\sigma +1}}{\rho _{0}^{\sigma }}  \notag \\
&\times &(1-x\delta ^{2})+\frac{1}{\rho _{0}}\sum_{\tau ,\tau ^{\prime
}}C_{\tau ,\tau ^{\prime }}  \notag \\
&\times &\int \int d^{3}pd^{3}p^{\prime }\frac{f_{\tau }(\vec{r},\vec{p}%
)f_{\tau ^{\prime }}(\vec{r},\vec{p}^{\prime
})}{1+(\vec{p}-\vec{p}^{\prime })^{2}/\Lambda ^{2}}. \label{MDIV}
\end{eqnarray}%
In the mean-field approximation, Eq.~(\ref{MDIV}) leads to the
following single-particle potential~\cite{Das03,Che05}
\begin{eqnarray}
U_\tau(\rho ,\delta ,\vec{p}) &=&A_{u}\frac{\rho _{-\tau }}{\rho _{0}}%
+A_{l}\frac{\rho _{\tau }}{\rho _{0}}  \notag \\
&+&B\left(\frac{\rho }{\rho _{0}}\right)^{\sigma }(1-x\delta ^{2})-4\tau x\frac{B}{%
\sigma +1}\frac{\rho ^{\sigma -1}}{\rho _{0}^{\sigma }}\delta \rho _{-\tau }
\notag \\
&+&\frac{2C_l}{\rho _{0}}\int d^{3}p^{\prime }\frac{f_{\tau }(%
\vec{r},\vec{p}^{\prime })}{1+(\vec{p}-\vec{p}^{\prime })^{2}/\Lambda ^{2}}
\notag \\
&+&\frac{2C_u}{\rho _{0}}\int d^{3}p^{\prime }\frac{f_{-\tau }(%
\vec{r},\vec{p}^{\prime })}{1+(\vec{p}-\vec{p}^{\prime })^{2}/\Lambda ^{2}}.
\label{MDIU}
\end{eqnarray}%
In the above, $\rho_n$ and $\rho_p$ are number densities of neutrons
and protons, respectively, and the isospin asymmetry $\delta$ is
defined as $\delta=(\rho_n-\rho_p)/\rho$, with $\rho=\rho_n+\rho_p$
being the total number density. $f_{\tau }(\vec{r},\vec{p})$ is the
phase-space distribution function, with $\tau=1(-1)$ for neutrons
(protons) being the isospin index.

The seven parameters ($A_l$, $A_u$, $B$, $C_l=C_{\tau,\tau}$,
$C_u=C_{\tau,-\tau}$, $\Lambda$, $\sigma$) can be fitted by seven
empirical constraints. Typically, five isoscalar constraints of the
saturation density $\rho_0$, the binding energy $E_0$, the
incompressibility $K_0$, the isoscalar effective mass $m_s^\star$,
and the single-particle potential $U_{0,\infty}$ at infinitely large
nucleon momentum at saturation density in symmetric nuclear matter
can be determined by $A_l+A_u$, $B$, $C_l+C_u$, $\Lambda$, and
$\sigma$. In addition, two isovector constraints of the symmetry
energy $E_{sym}(\rho_0)$ and the symmetry potential $U_{sym,\infty}$
at infinitely large nucleon momentum (or equivalently the
neutron-proton effective mass splitting) at saturation density can
be determined by $A_l-A_u$ and $C_l-C_u$. In addition to the $x$
parameter in the previous MDI interaction which can be used to
adjust the slope parameter $L$ of the symmetry energy at saturation
density, we introduce two addition parameters $y$ and $z$ to adjust
respectively $U_{sym,\infty}$ and $E_{sym}(\rho_0)$, and $A_l$,
$A_u$, $C_l$, and $C_u$ can then be expressed as
\begin{eqnarray}
A_{l}(x,y)&=&A_{l0} + y + x\frac{2B}{\sigma +1},  \label{AlImMDI}\\
A_{u}(x,y)&=&A_{u0} - y - x\frac{2B}{\sigma +1}, \label{AuImMDI}\\
C_{l}(y,z)&=&C_{l0} - 2(y-2z)\frac{p^2_{f0}}{\Lambda^2\ln [(4 p^2_{f0} + \Lambda^2)/\Lambda^2]}, \label{ClImMDI}\\
C_{u}(y,z)&=&C_{u0} + 2(y-2z)\frac{p^2_{f0}}{\Lambda^2\ln [(4 p^2_{f0} + \Lambda^2)/\Lambda^2]}, \label{CuImMDI}
\end{eqnarray}
where $p_{f0}$ is the nucleon Fermi momentum in symmetric nuclear
matter at saturation density. For $x=0$, $y=0$, and $z=0$, we choose
the following empirical values, i.e., $\rho_0 = 0.16$ fm$^{-3}$,
$E_0(\rho_0) = -16$ MeV, $K_0 = 230$ MeV, $m^\star_s = 0.7 m$,
$E_{sym}(\rho_0) = 32.5$ MeV, and $U_{0,\infty} = 75$ MeV, which
lead to $A_{l0} = A_{u0}= -66.963$ MeV, $B = 141.963$ MeV, $C_{l0} =
-60.4860$ MeV, $C_{u0} =  -99.7017$ MeV, $\Lambda = 2.42401p_{f0}$,
and $\sigma = 1.26521$. Again, the values of $x$, $y$, and $z$ will
only affect the isovector properties of nuclear matter but will not
lead to the variation of the empirical isoscalar constraints.

The potential energy density functional of Eq.~(\ref{MDIV}) can be
obtained from the following effective NN interaction within
Hartree-Fock approach~\cite{Das03,DME}
\begin{eqnarray}
v(\vec{r}_1,\vec{r}_2) &=& \frac{1}{6}t_3(1+x_3 P_\sigma)
\rho^\gamma\left(\frac{\vec{r}_1+\vec{r}_2}{2}\right)
\delta(\vec{r}_1-\vec{r}_2) \notag\\
&+& (W+G P_\sigma - H P_\tau - M P_\sigma P_\tau) \frac{e^{-\mu
|\vec{r}_1-\vec{r}_2}|}{|\vec{r}_1-\vec{r}_2|},\notag\\
\label{MDIyuk}
\end{eqnarray}
namely, a density-dependent zero-range interaction and a
finite-range Yukawa-type two-body interaction, with $\vec{r}_1$ and
$\vec{r}_2$ being the spatial coordinates of the two nucleons and
$P_\sigma$ and $P_\tau$ being the spin and isospin exchange
operator. The values of the parameters $t_3$, $\gamma$, $W$, $G$,
$H$, $M$, and $\mu$ can be uniquely determined from $A_l$, $A_u$,
$B$, $C_l$, $C_u$, $\Lambda$, and $\sigma$~\cite{DME}. The $x$
parameter is related to the value of $x_3$, i.e., the relative
contribution of the isospin-singlet and the isospin-triplet channel
of the density-dependent interaction, while the values of $y$ and
$z$ are related to those of $W$, $G$, $H$, and $M$ and are thus
determined by the different spin-isospin channels of the
finite-range interaction.

\begin{figure}[htb]
\centering
\includegraphics[scale=0.8]{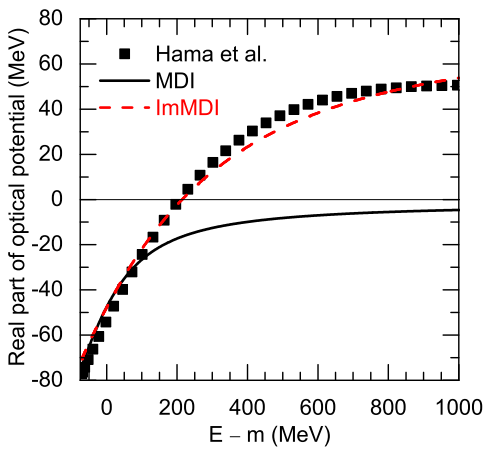}
\caption{(Color online) The ImMDI interaction prediction on the
single-particle potential in symmetric nuclear matter at $\rho_0$ as
a function of nucleon total energy subtracted by its rest mass. The
results of the MDI interaction and the optical potential by Hama
{\it et al.}~\cite{Hama90,Coo93} are also shown for comparison.}
\label{HamaImMDI}
\end{figure}

In the ImMDI interaction, $U_{0,\infty} = (A_l+A_u)/2+B = 75$ MeV is
selected to fit the empirical optical potential by Hama {\it et
al.}, and this can be seen from Fig.~\ref{HamaImMDI} where the
single-particle potential (real part of optical potential) in
symmetric nuclear matter at $\rho_0$ is plotted as a function of
nucleon total energy subtracted by its rest mass, i.e., $E-m$. The
results of the MDI interaction and the optical potential by Hama
{\it et al.}~\cite{Hama90,Coo93} are also shown for comparison. One
can see that the MDI interaction, whose momentum dependence of the
mean-field potential is fitted to reproduce that of the Gogny
interaction, significantly under-predicts the empirical optical
potential by Hama {\it et al.} when $E-m$ is larger than about $160$
MeV. We note that the wrong asymptotic value of the isoscalar
potential at high momentum is actually a longstanding problem of the
Gogny effective interaction. On the other hand, the energy/momentum
dependence of the single-particle potential in symmetric nuclear
matter at $\rho_0$ predicted by the ImMDI interaction is in good
agreement with the empirical optical potential by Hama {\it et al.}
in the whole energy region up to about $E-m = 1000$ MeV. Therefore,
the ImMDI interaction provides a reasonable choice for the transport
model simulations for heavy-ion collisions at low and intermediate
energies (up to at least about 1 GeV/nucleon).

\begin{figure}[ht]
\centerline{\includegraphics[scale=0.7]{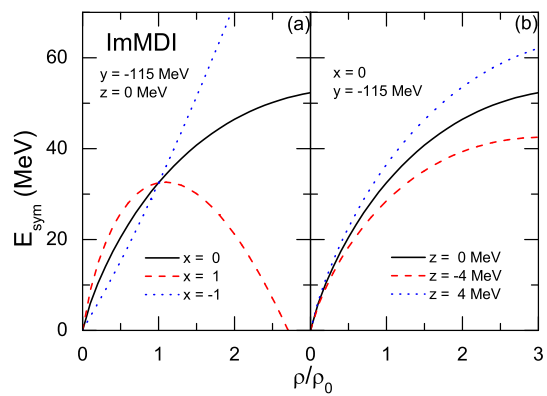}} \caption{(Color
online) The symmetry energy from the ImMDI interaction by adjusting
the value of parameter $x$ at $y=-115$ MeV and $z=0$ MeV (a) or
parameter $z$ at $x=0$ and $y=-115$ MeV (b).} \label{xz}
\end{figure}

\begin{figure}[htb]
\centering
\includegraphics[scale=0.7]{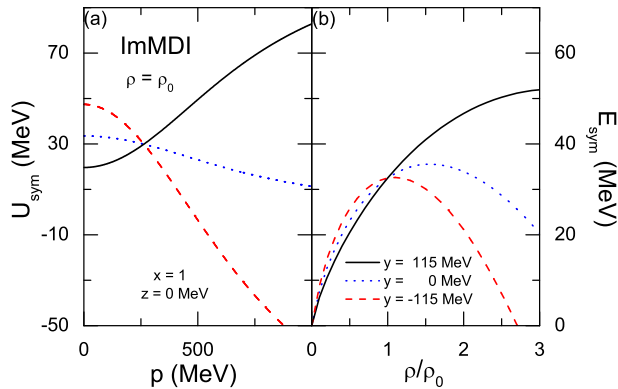}
\caption{(Color online) The symmetry potential at saturation density
(a) and symmetry energy (b) from the ImMDI interaction by adjusting
the value of parameter $y$ at $x=1$ and $z=0$ MeV.} \label{y}
\end{figure}

In the ImMDI interaction, one can vary flexibly three parameters,
i.e., $x$, $y$, and $z$ to change the isovector properties of
nuclear matter. Similar to the previous MDI interaction, the density
dependence of the symmetry energy (e.g., the slope parameter $L$)
changes with the parameters $x$ while $E_{sym}(\rho_0)$ remains
unchanged, as can be seen from the left panel of Fig.~\ref{xz}. On
the other hand, the value of the symmetry energy at saturation
density changes from $E_{sym}(\rho_0)$ to $E_{sym}(\rho_0)+z$ when
$z$ is adjusted, as can be seen from the right panel of
Fig.~\ref{xz}. In this way one can easily study the sensitivity of
the isospin tracers to the values of $L$ and $E_{sym}(\rho_0)$
simultaneously. In addition, one can vary the $y$ parameter, which
is equivalent to $U_{{sym},\infty}$, to modify the momentum
dependence of the symmetry potential $U_{sym}(\rho,p)$ at $\rho_0$
(and also other densities), while in the MDI interaction, the
momentum dependence of $U_{sym}(\rho,p)$ is fixed although the
magnitude of $U_{sym}(\rho,p)$ at non-saturation densities can be
varied using different $x$ values. It is clearly seen from the left
panel of Fig.~\ref{y} that one can flexibly vary $y$ parameter to
mimic different momentum/energy dependences of the $U_{sym}(\rho
,p)$ (and thus isospin splitting of nucleon effective mass),
providing a convenient way to explore the consequent effects in
heavy-ion collisions. In addition, one can see $U_{sym}(\rho_0 ,p)$
at $p = p_{f0}$ (corresponding to a nucleon kinetic energy of $36.8$
MeV) is independent of the $y$ parameter by construction. On the
other hand, it is seen from the right panel of Fig.~\ref{y} that the
density dependence of the symmetry energy changes with $y$ as well,
with the values of $E_{sym}(\rho_0)$ fixed. This can be understood
as the slope parameter $L$ depends on not only the magnitude of
symmetry potential, which is related to the $x$ parameter, but also
the momentum dependence of the symmetry
potential~\cite{XuC10,ChR12}.

\section{Effects of neutron-proton effective mass splitting}

The ImMDI interaction described in the previous section provides
possibilities of studying more detailed isovector properties of
nuclear matter flexibly. In the following, we study the effects of
neutron-proton effective mass splitting on thermodynamical
properties of neutron-rich nuclear matter. One can see from
Figs.~\ref{xz} and \ref{y} that [($x=0$), ($y=-115$ MeV)] and
[($x=1$), ($y=115$ MeV)] give almost the same density dependence of
the symmetry energy at $z=0$, while the two parameter sets lead to
two extreme momentum dependences of the symmetry potential, with
$U_{sym}$ from [($x=0$), ($y=-115$ MeV)] decreases with increasing
nucleon momentum and thus $m_n^\star>m_p^\star$ and that from
[($x=1$), ($y=115$ MeV)] increases with increasing nucleon momentum
and thus $m_n^\star<m_p^\star$. We will carry out our study based on
the two parameter sets in the following.

\subsection{temperature dependence of symmetry energy}

\begin{figure}[h]
\centerline{\includegraphics[scale=0.7]{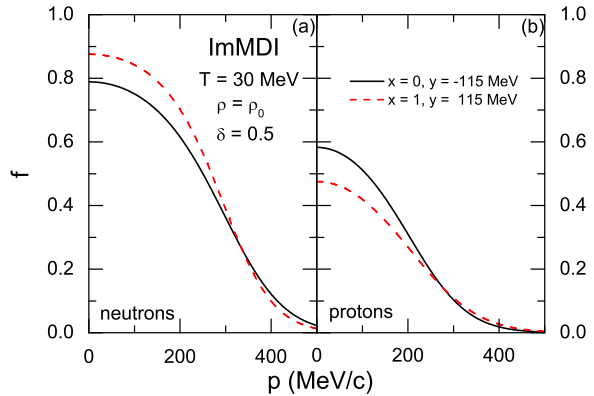}} \caption{(Color
online) The equilibrated phase-space distribution functions
(normalized by the spin degeneracy) of neutrons (a) and protons (b)
from the ImMDI interaction for [($x=0$), ($y=-115$ MeV)] and
[($x=1$), ($y=115$ MeV)] in neutron-rich nuclear matter of isospin
asymmetry $\delta=0.5$ at saturation density and temperature $T=30$
MeV.} \label{f}
\end{figure}

Since from Eq.~(\ref{MDIU}) the single-particle potential depends on
the phase-space distribution function, and from single particle
approximation this Fermi-Dirac phase-space distribution function in
equilibrium depends on the single-particle potential, an iteration
method is needed to calculate the mean-field potential and the
equation of state at finite temperatures~\cite{Xu07a}. From such a
self-consistent calculation, the equilibrated phase-space
distribution functions of neutrons and protons for [($x=0$),
($y=-115$ MeV)] and [($x=1$), ($y=115$ MeV)] in neutron-rich nuclear
matter of isospin asymmetry $\delta=0.5$ at saturation density and
temperature $T=30$ MeV are displayed in Fig.~\ref{f}. It is seen
that [($x=0$), ($y=-115$ MeV)], giving a larger neutron effective
mass than proton, has a more diffusive distribution for neutrons and
less diffusive distribution for protons compared to [($x=1$),
($y=115$ MeV)]. This is understandable as the self-consistent
calculation balances the energy of the system at fixed isospin
asymmetry, so for a larger neutron (proton) effective mass than
proton (neutron) with [($x=0$), ($y=-115$ MeV)] ([($x=1$), ($y=115$
MeV)]) more neutrons (protons) are allowed to occupy the
high-momentum states.

\begin{figure}[h]
\centerline{\includegraphics[scale=0.7]{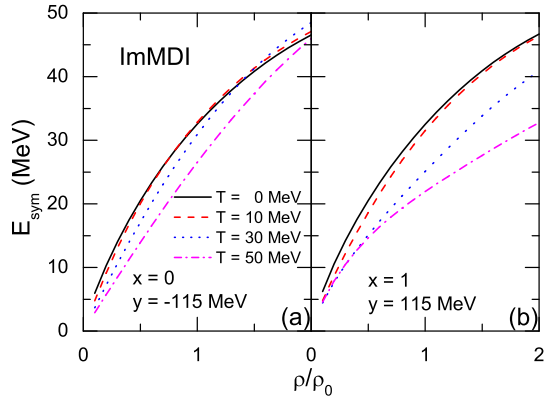}} \caption{(Color
online) The nuclear symmetry energy from the ImMDI interaction for
[($x=0$), ($y=-115$ MeV)] (a) and [($x=1$), ($y=115$ MeV)] (b) at
temperatures of $0$, $10$, $30$, and $50$ MeV.} \label{EsymT}
\end{figure}

\begin{figure}[h]
\centerline{\includegraphics[scale=0.7]{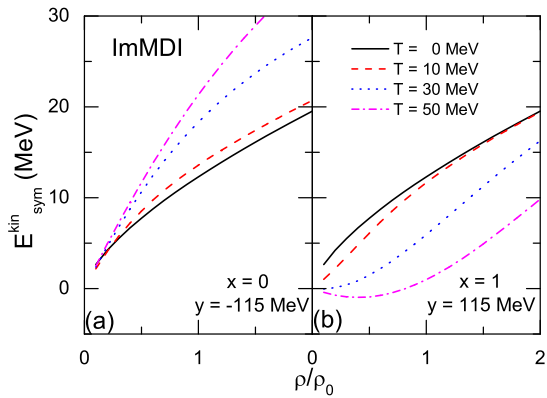}}
\caption{(Color online) Same as Fig.~\ref{EsymT} but only for the
kinetic contribution of the symmetry energy.} \label{EsymkinT}
\end{figure}

\begin{figure}[h]
\centerline{\includegraphics[scale=0.7]{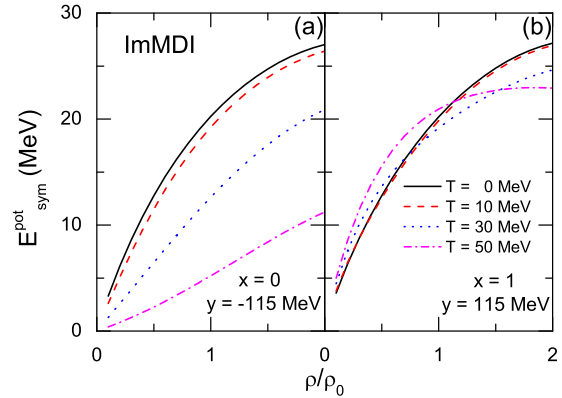}}
\caption{(Color online) Same as Fig.~\ref{EsymT} but only for the
potential contribution of the symmetry energy.} \label{EsympotT}
\end{figure}

As a key quantity of isospin physics, the density dependence of the
symmetry energy for [($x=0$), ($y=-115$ MeV)] and [($x=1$), ($y=115$
MeV)] at different temperatures are shown in Fig.~\ref{EsymT}. At
finite temperatures the symmetry energy is calculated numerically by
taking the difference of the binding energy at $\delta=0$ and
$\delta=0.2$. One can see for [($x=0$), ($y=-115$ MeV)] the symmetry
energy decreases with increasing temperature at lower densities but
slightly increases with increasing temperature at higher densities,
while for [($x=1$), ($y=115$ MeV)] the symmetry energy decreases
with increasing temperature at all the densities. Similar behavior
was observed in Ref.~\cite{OLi11} based on the Skyrme-Hartree-Fock
functional. To understand the different temperature dependence of
the symmetry energy with different isospin splitting of nucleon
effective mass, we further show in Figs.~\ref{EsymkinT} and
\ref{EsympotT} the kinetic and potential contribution to the
symmetry energy, respectively. It is interesting to see that the
kinetic contribution to the symmetry energy increases with
increasing temperature for [($x=0$), ($y=-115$ MeV)] but decreases
with increasing temperature for [($x=1$), ($y=115$ MeV)]. This is
because there are more neutrons and less protons in the high-energy
states with increasing temperature for [($x=0$), ($y=-115$ MeV)] but
it is opposite for [($x=1$), ($y=115$ MeV)], as can be seen from
Fig.~\ref{f}. For the potential contribution to the symmetry energy,
it somehow decreases with increasing temperature for [($x=0$),
($y=-115$ MeV)] but has a weak temperature dependence for [($x=1$),
($y=115$ MeV)]. The combination of Figs.~\ref{EsymkinT} and
\ref{EsympotT} leads to the temperature dependence of the total
symmetry energy in Fig.~\ref{EsymT}.

\subsection{isovector single-particle properties}

\begin{figure}[h]
\centerline{\includegraphics[scale=0.7]{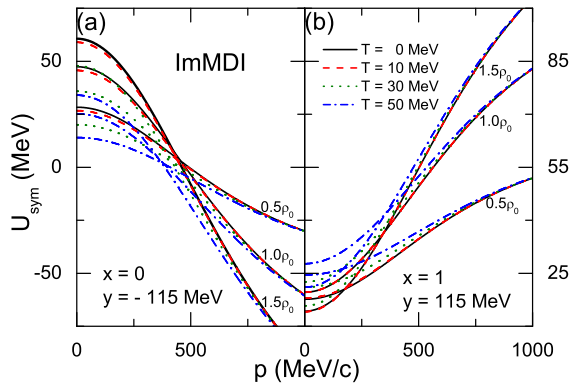}} \caption{(Color
online) The momentum dependence of the symmetry potential from the
ImMDI interaction for [($x=0$), ($y=-115$ MeV)] (a) and [($x=1$),
($y=115$ MeV)] (b) at different densities and temperatures. }
\label{Usym}
\end{figure}

We now move to the isovector single-particle properties of nuclear
matter including the symmetry potential and the neutron-proton
effective mass splitting. The momentum dependence of the symmetry
potential for [($x=0$), ($y=-115$ MeV)] and [($x=1$), ($y=115$ MeV)]
at different densities and temperatures are shown in
Fig.~\ref{Usym}, and the results are calculated by taking the
potential difference of neutrons and protons at $\delta=0.2$. One
can see that the symmetry potential decreases with increasing
momentum for [($x=0$), ($y=-115$ MeV)] but increases with increasing
momentum for [($x=1$), ($y=115$ MeV)], and the slope is larger at
higher densities. The symmetry potential becomes negative at high
nucleon momenta for [($x=0$), ($y=-115$ MeV)] while it is always
positive for [($x=1$), ($y=115$ MeV)]. With the increasing
temperature, only the low-momentum part of the symmetry potential is
affected while the high-momentum part remains almost unchanged. It
is interesting to see that symmetry potential decreases with
increasing temperature for [($x=0$), ($y=-115$ MeV)] while it
increases with increasing temperature for [($x=1$), ($y=115$ MeV)].

\begin{figure}[h]
\centerline{\includegraphics[scale=0.7]{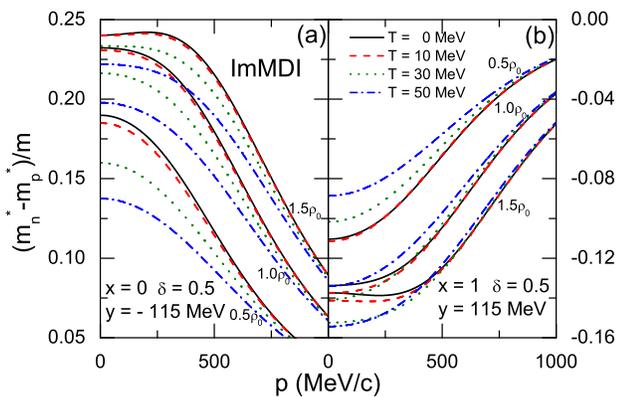}} \caption{(Color
online) The relative neutron-proton effective mass splitting from
the ImMDI interaction for [($x=0$), ($y=-115$ MeV)] (a) and
[($x=1$), ($y=115$ MeV)] (b) in neutron-rich nuclear matter of
isospin asymmetry $\delta=0.5$ at different densities and
temperatures.} \label{mstar}
\end{figure}

A positive symmetry potential gives repulsive force to neutrons and
attractive force to protons, while the velocity of the nucleon
depends not only on the force but also on the in-medium effective
mass. The nucleon effective mass, which is defined as
\begin{equation}
\frac{m_{\tau }^{\ast }}{m}=\left( 1+\frac{m}{p}\frac{dU_{\tau
}}{dp}\right) ^{-1}, \label{Meff}
\end{equation}%
is a function of nucleon momentum but mostly represented by the
value at Fermi momentum. The relative neutron-proton effective mass
splitting for [($x=0$), ($y=-115$ MeV)] and [($x=1$), ($y=115$ MeV)]
in neutron-rich nuclear matter of isospin asymmetry $\delta=0.5$ at
different densities and temperatures are shown in Fig.~\ref{mstar}.
Indeed, the neutron effective mass is larger than protons for
[($x=0$), ($y=-115$ MeV)] and smaller than protons for [($x=1$),
($y=115$ MeV)] at all the densities and temperatures. Generally, the
relative effective mass splitting is smaller at higher nucleon
momenta and stronger at higher densities, and the splitting becomes
weaker at higher temperatures for [($x=0$), ($y=-115$ MeV)] but the
temperature dependence is somehow complicated for [($x=1$), ($y=115$
MeV)] especially at higher densities.

\subsection{differential isospin fractionation}

The two phases of nuclear matter can coexist if the Gibbs condition
is satisfied, i.e., they have the same temperature, pressure, and
chemical potential. The dense phase with smaller isospin asymmetry
is called the liquid phase, while the dilute phase with larger
isospin asymmetry is called the gas phase. As the symmetry energy
generally increases with increasing density at least at
subsaturation densities, the high-density phase should have a
smaller isospin asymmetry while the low-density phase can have a
larger isospin asymmetry, so in this way the total energy can be
well distributed in the two phases and reach a minimum value. This
is the so-called isospin fractionation.

\begin{figure}[h]
\centerline{\includegraphics[scale=0.7]{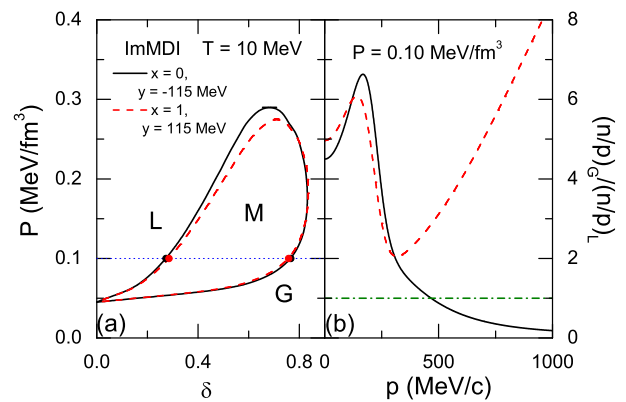}}
\caption{(Color online) Left panel: The section of binodal surface
from the ImMDI interaction for [($x=0$), ($y=-115$ MeV)] and
[($x=1$), ($y=115$ MeV)] at a temperature $T=10$ MeV, and $L$, $G$,
and $M$ represent the liquid phase, the gas phase, and the mixed
phase, respectively; Right panel: The double neutron/proton ratio in
gas and liquid phases $(n/p)_{G}/(n/p)_{L}$ at the pressure of
$0.10$ MeV/fm$^3$ as a function of nucleon momentum.}
\label{isofrac}
\end{figure}

Numerically, the binodal surface of nuclear liquid-gas phase
transition can be constructed by drawing rectangles in the chemical
potential isobars of neutrons and protons as functions of isospin
asymmetry at a given temperature~\cite{Mul95,Xu07b}. The obtained
two phases thus satisfy the Gibbs condition, with the one of larger
isospin asymmetry corresponding to the gas phase and that of the
smaller isospin asymmetry corresponding to the liquid phase.
Collecting all such pairs at each pressure forms the binodal surface
of the nuclear liquid-gas phase transition, as shown in the left
panel of Fig.~\ref{isofrac} at temperature $T=10$ MeV. The binodal
surface is useful in calculating the volume fraction of each phase
and studying the properties of nuclear liquid-gas phase transition
at fixed isospin asymmetry as shown in Ref.~\cite{Xu08}, and the
liquid phase (L), the gas phase (G), and the mixed phase (M) are
denoted in the figure. One can see that the binodal surface is
similar for [($x=0$), ($y=-115$ MeV)] and [($x=1$), ($y=115$ MeV)].
This is not surprising as both the chemical potential and the
pressure are determined from the equation of state, which is almost
the same for the two parameter sets. The slight difference is
expected to be due to the different temperature dependence of the
symmetry energy.

Similar to Ref.~\cite{Li07}, we study the differential isospin
fractionation at pressure $P=0.1$ MeV/fm$^3$. As can be seen from
the left panel of Fig.~\ref{isofrac}, the nuclear matter in the
mixed phase region at $P=0.1$ MeV/fm$^3$ composes of the liquid
phase and the gas phase at two edges of the binodal surface with the
same pressure. For [($x=0$), ($y=-115$ MeV)], the densities and
isospin asymmetries of the liquid and gas phases are
$\rho_L=0.757\rho_0$, $\delta_L=0.273$, $\rho_G=0.087\rho_0$, and
$\delta_G=0.766$, respectively. For [($x=1$), ($y=115$ MeV)], the
densities and isospin asymmetries of the liquid and gas phases are
$\rho_L=0.752\rho_0$, $\delta_L=0.285$, $\rho_G=0.083\rho_0$, and
$\delta_G=0.758$, respectively. Thus, the ratios of neutron/proton
in the gas phase to that in the liquid phase, i.e,
$(n/p)_G/(n/p)_L$, are 4.31 for [($x=0$), ($y=-115$ MeV)] and 4.04
for [($x=1$), ($y=115$ MeV)]. Although the total ratios are similar
for the two parameter sets, the differential behaviors, i.e., the
momentum dependence, are quite different, as can be seen from the
right panel of Fig.~\ref{isofrac}. Similar to the findings in
Ref.~\cite{Li07}, the $(n/p)_G/(n/p)_L$ ratio becomes smaller than 1
when the nucleon momentum is larger than about $500$ MeV/c. This can
be understood by checking with the symmetry potential in
Fig.~\ref{Usym} that $U_{sym}$ becomes negative when the nucleon
momentum is larger than about $500$ MeV/c. For [($x=1$), ($y=115$
MeV)], since the symmetry potential is always positive and is larger
at higher nucleon momenta, the $(n/p)_G/(n/p)_L$ ratio is always
larger than 1 and increases with increasing momentum at higher
nucleon energies. In intermediate-energy heavy-ion collisions, the
gas phase is formed by free nucleons while the liquid phase is
formed by those in heavy clusters. Consistent with the finding here,
it was show in Refs.~\cite{LiBA04,Riz05,Fen12,Zha14} that the
neutron/proton ratio of energetic nucleons is sensitive to the
neutron-proton effective mass splitting.

\section{summary}

Based on an improved isospin- and momentum-dependent interaction,
with the isoscalar single-nucleon potential refitted to that
extracted by optical model analyses of proton-nucleus scattering
data up to nucleon kinetic energy of about 1 GeV/c, and three parameters
included for studing the detailed isovector properties of nuclear
matter, i.e., the slope parameter of the symmetry energy, the
momentum dependence of the symmetry potential, and the symmetry
energy at saturation density, we have studied the thermodynamical
properties of neutron-rich nuclear matter with the same equation of
state but different neutron-proton effective mass splittings. We
found that the phase-space distribution in equilibrium, the
temperature dependence of the symmetry energy, and the differential
isospin fractionation can be affected by the isospin splitting of
nucleon effective mass.

\section{acknowledgments}

This work was supported in part by the Major State Basic Research
Development Program (973 program) in China under Contract Nos.
2015CB856904, 2014CB845401, and 2013CB834405, the National Natural
Science Foundation of China under Grant Nos. 11475243, 11275125,
11135011, and 11320101004, the "100-talent plan" of Shanghai
Institute of Applied Physics under Grant No. Y290061011 from the
Chinese Academy of Sciences, the "Shanghai Pujiang Program" under
Grant No. 13PJ1410600, the ¡°Shu Guang¡± project supported by
Shanghai Municipal Education Commission and Shanghai Education
Development Foundation, the Program for Professor of Special
Appointment (Eastern Scholar) at Shanghai Institutions of Higher
Learning, the Science and Technology Commission of Shanghai
Municipality (11DZ2260700), the US National Science Foundation
grants PHY-1068022, the National Aeronautics and Space
Administration under grant NNX11AC41G issued through the Science
Mission Directorate, and the CUSTIPEN (China-U.S. Theory Institute
for Physics with Exotic Nuclei) under DOE grant number
DE-FG02-13ER42025.

\end{document}